\documentclass[lettersize,journal]{IEEEtran}
\usepackage{amsfonts}
\usepackage{array, multirow, tabularx}
\newcolumntype{C}[1]{>{\centering\arraybackslash}p{#1}} 
\newcolumntype{Y}{>{\centering\arraybackslash}X}       
\usepackage{algorithm}
\usepackage{algpseudocode}
\usepackage{amsmath,amssymb}
\usepackage{graphicx}
\usepackage[caption=false,font=footnotesize]{subfig}
\usepackage{textcomp}
\usepackage{stfloats}
\usepackage{url}
\usepackage{verbatim}
\usepackage{cite}
\usepackage{xcolor}
\newcommand{\rev}[1]{\textcolor{black}{#1}}
\usepackage{placeins}

\begin{document}

\title{FedAvgIm: Federated Learning for Multi-Site Impedance Identification of Inverter-Based Resources}

\author{Quang Manh Hoang
\IEEEmembership{Member, IEEE}, Guilherme Vieira Hollweg \IEEEmembership{Member, IEEE}, Akhtar Hussain \IEEEmembership{Member, IEEE}, Wencong Su \IEEEmembership{Senior Member, IEEE}, Van-Hai Bui \IEEEmembership{Senior Member, IEEE}
\thanks{Quang Manh Hoang,  Guilherme Vieira Hollweg, Wencong Su, and Van-Hai Bui are with Department of Electrical and Computer Engineering, University of Michigan-Dearborn, Michigan, USA (e-mail: manhhq@umich.edu; hollweg@umich.edu; wencong@umich.edu).
Corresponding author: Van-Hai Bui (e-mail: vhbui@umich.edu).}
\thanks{ Akhtar Hussain is with Department of Electrical and Computer Engineering, Laval University, Quebec City, Canada (e-mail: akhtar.hussain@gel.ulaval.ca).}
}

\IEEEaftertitletext{%
  \vspace{0.5\baselineskip}
  \begin{minipage}{\textwidth}
    \footnotesize
    \noindent
    \copyright~2026 IEEE. Personal use of this material is permitted.  Permission from IEEE must be obtained for all other uses, in any current or future media, including reprinting/republishing this material for advertising or promotional purposes, creating new collective works, for resale or redistribution to servers or lists, or reuse of any copyrighted component of this work in other works.
    \par\smallskip
    \footnotesize
    \noindent
    \textit{This is the author-accepted version
    of an article published in IEEE Transactions
    on Smart Grid. The final published version
    is available at:}
    \url{https://doi.org/10.1109/TSG.2026.3717637}
  \end{minipage}%
}


\maketitle

\begin{abstract}
With the increasing integration of inverter-based resources (IBRs) into power systems, assessing small-signal interaction stability has become essential. Impedance/admittance models enable frequency-domain characterization of IBR dynamics without requiring access to proprietary internal control details, but measurement-based identification is time-consuming and must be repeated across operating points. This paper proposes FedAvgIm, a privacy-aware federated-learning framework for multi-site dq-admittance identification under severe local data scarcity and raw-data-locality constraints. Each IBR site trains a local feed-forward neural network using a small set of operating-point--frequency samples, while only model parameters are shared with a cloud aggregator that performs sample-size-weighted FedAvg to update a global model and broadcast it back to all sites. To demonstrate effectiveness, EMT measurement-based admittance datasets are constructed for nine grid-following IBRs with different parameters, architectures, measurement-noise levels, and non-IID settings, where each client has only 40--390 training samples. Results show that FedAvgIm achieves performance close to centralized training and consistently outperforms local-only learning and raw-data-locality-preserving transfer-learning baselines. Additional studies show that FedAvgIm maintains stable learning behavior under measurement noise and remains beneficial under moderate non-IID conditions. The estimated dq-admittance models are further applied to generalized Nyquist stability analysis and validated against EMT time-domain simulations under multiple operating points and grid-strength conditions.
\end{abstract}

\begin{IEEEkeywords}
Decentralized learning, data locality, dq-admittance identification, impedance-based stability analysis, inverter-based resources.
\end{IEEEkeywords}

\section{Introduction}
\IEEEPARstart{M}{odern} power grids are expected to integrate an increasing number of inverter-based resources (IBRs), which makes the system more vulnerable and reduces the applicability of traditional control and stability analysis methods. One of the major challenges is the complex interaction among multiple IBRs due to their wide range of control bandwidths, topologies, and algorithms, together with the reduced inertia compared to traditional synchronous-generator (SGs)-based power systems. These interactions can lead to small-signal stability problems such as sub-synchronous oscillation, near-synchronous oscillation, and super-synchronous resonance \cite{shah2019impedance}.

In traditional power systems dominated by SGs, small-signal stability can be studied using white-box state-space models because SGs have standardized structures and well-known parameters. Consequently, eigenvalue analysis can be directly applied to assess system stability \cite{kundur2007power}. However, this approach is often not applicable to inverter-dominated power systems, where the control structures and parameters of IBRs are frequently kept confidential by power-converter manufacturers for security and intellectual-property reasons \cite{wu2023survey}. As a result, impedance-based stability analysis has emerged as a powerful frequency-domain tool for assessing the small-signal stability of black-box systems \cite{shah2019impedance, zhou2022comprehensive, li2024machine, hadjileonidas2025application}.

Impedance-based analysis is commonly formulated in three largely equivalent domains: dq impedance, sequence impedance, and phasor impedance \cite{shah2019impedance}. The dq formulation represents terminal behavior as a $2\times2$ multiple-input multiple-output (MIMO) impedance/admittance matrix in the synchronous rotating frame, thereby preserving control-induced couplings and enabling a general, mechanism-revealing stability assessment. By contrast, the sequence formulation is often appealing because it offers a single-input single-output (SISO)-like interpretation and facilitates intuitive Nyquist/Bode analysis. However, recent studies have highlighted frequency-coupling effects, particularly at low frequencies, that can couple positive and negative-sequence dynamics and undermine the validity of decoupled SISO assumptions. Phasor impedance has been introduced to better align with grid-level phasor-network tools and improve scalability, but it is less direct for diagnosing dq-channel coupling mechanisms. Therefore, this work adopts the dq-impedance framework to avoid potentially restrictive decoupling assumptions and to consistently capture both IBR--grid and IBR--IBR interactions over the frequency range of interest.

Regardless of the specific hardware (e.g., LCL-filter parameters and topology) and software (e.g., control algorithms), the dynamic characteristics of IBRs can be represented using impedance/admittance models, which describe the small-signal current/voltage responses to perturbations around an operating point. There are several ways to obtain IBR admittance/impedance models. One approach is analytical modeling, which linearizes the IBR dynamics around a steady-state operating condition \cite{wen2015analysis, guo2021impedance, zhou2022comprehensive, xu2023small}. This approach is useful for studying IBR dynamics under various conditions. For example, \cite{wen2015analysis} derives a dq-domain impedance model of a grid-following inverter and shows that the phase-locked loop (PLL) bandwidth and operating point (active/reactive power setpoints) mainly affect the $q$-axis impedance and, under certain conditions, may lead to instability.

However, the formulation of small-signal admittance models often depends on modeling choices and engineering experience, and it may need to be revised when the IBR structure changes. To alleviate this limitation, the component-connection method, originally introduced for dynamic studies of conventional AC/DC systems \cite{gaba2002comparative}, has recently been adopted as a systematic way to derive IBR admittance models \cite{zhou2022comprehensive, xu2023small}. In this method, each subsystem component (e.g., PLL, computational delay, current controller, and filter) is represented in a standard state-space form. Interconnection matrices are then used to couple the subsystem blocks, yielding an overall state-space model from which the admittance can be computed. Nevertheless, analytical approaches remain most suitable when the IBR architecture and parameters are available. In practice, this assumption often does not hold due to limited standardization and strong commercial incentives, vendors typically treat control algorithms and parameters as proprietary and may only provide black-box input-output models without internal details.

In practice, impedance/admittance models are typically obtained using measurement-based methods, in which small current or voltage perturbations are injected at selected frequencies and the corresponding responses are measured \cite{francis2011algorithm, 9137400}. For impedance/admittance measurement and EMT-based identification, the dominant approach is frequency scanning \cite{das2025survey}. The system is first driven to steady state then a small, known perturbation (voltage or current) is injected. Terminal voltages/currents are recorded, and a fast Fourier transform (FFT) is used to extract the phasors at the injection frequency, from which the admittance at that frequency is computed. Repeating this procedure over a set of frequencies yields the full admittance profile. Scans can be performed in the positive-sequence domain (SISO, typically requiring one scan) or in the $dq/\alpha\beta$ domain (MIMO, requiring separate axis excitations). Perturbations may be single-tone or multi-tone and can take various shapes (step, ramp, chirp, or impulse) \cite{das2025survey}. The main practical drawback is that frequency scanning is time-consuming and computationally intensive. Single-tone scanning requires one perturbation per frequency point, and MIMO scanning further increases the effort because each axis must be excited separately. Even for a single IBR with a fully known control structure, obtaining wide-band admittance characteristics across multiple operating points can take several hours. In inverter-dominated systems with highly variable operating conditions, many such scenarios must be analyzed, which further increases the overall computational burden and complexity.

To overcome these drawbacks, several studies have leveraged data-driven methods to identify black-box IBR impedance/admittance models across denser operating conditions using only a limited number of measurement samples \cite{zhang2020artificial, mohammed2024support, li2024machine, xiao2009recurrent, lyu2023data, zhang2022physics, zhang2023transfer, rane2025transfer}. An early effort to use data-driven techniques for wide-frequency impedance identification of power-electronic converters was reported in 2009 \cite{xiao2009recurrent}, however, it focused on a DC system and considered only a single operating point. In \cite{zhang2020artificial}, an artificial neural network was used to identify the impedance of a single grid-following inverter under different operating points. Nevertheless, neural-network-based approaches are typically data-hungry and require a large amount of training data. Transfer learning offers a promising way to reduce the data requirement at the target site \cite{zhang2023transfer, rane2025transfer, li2024machine}. However, it still relies on a medium-to-large dataset for the pretraining stage. More recently, physics-informed machine learning has been applied to impedance identification \cite{zhang2022physics}, but it usually requires partial knowledge of the IBR internal architecture, which is not always available in practice. In \cite{mohammed2024support}, the authors used a support vector machine to predict the impedance of both grid-following and grid-forming inverters, yet the approach still requires a moderate number of data points. Importantly, when data privacy is \textcolor{blue}{a concern} and each IBR site has an extremely small dataset, existing methods may suffer from overfitting. Moreover, most existing machine-learning approaches model one IBR at a time and therefore cannot effectively leverage data from multiple IBRs to mitigate severe data scarcity.

To address these issues, this paper proposes FedAvgIm, a decentralized federated-learning framework based on the FedAvg algorithm \cite{mcmahan2017communication}, for multi-site IBR admittance identification under (i) strict data-locality requirements, (ii) extremely limited local datasets, (iii) unbalanced data across sites, where each IBR has a different number of samples, and (iv) the presence of multiple IBRs within a system. Specifically, EMT measurement-based admittance datasets are constructed for nine IBRs with different architectures, parameters, measurement-noise levels, and degrees of non-independent and identically distributed (non-IID) data. These EMT measurement-based datasets are used to represent the black-box identification scenario in which internal model parameters are unavailable. To satisfy raw-data sharing constraints, the datasets are stored locally at each IBR site and are used for decentralized training without aggregating raw data in a central location. The proposed framework is compared against centralized training, local-only training, and transfer-learning baselines. Its effectiveness is further evaluated under different measurement-noise conditions and non-IID levels. In addition, the estimated admittance models are used for small-signal stability assessment via the Nyquist criterion under various operating points and grid conditions.

The proposed framework offers the following advantages:
\begin{enumerate}
  \item Improved label efficiency under data scarcity: Each local IBR requires only a small number of measured samples because the model can leverage knowledge learned from other IBR sites, thereby reducing measurement time and effort.
    \item Privacy-aware collaboration: Raw data remain stored locally at each site, while only the parameters of the feed-forward neural network (FNN) are shared and updated. This reduces centralized raw-data exposure and helps reduce cybersecurity and intellectual-property risks associated with raw-data pooling.
  \item Practical validation: Noisy measurement data, frequency-domain stability assessment, and EMT time-domain simulations demonstrate the applicability of the proposed framework.
\end{enumerate}
The remainder of this paper is organized as follows. Section II provides an overview of admittance modeling methods and discusses the limitations of existing methods. Section III describes the proposed FedAvgIm framework, from data collection and organization to the proposed algorithm. Section IV presents the performance evaluation through six different case studies. Finally, Section V concludes the paper.
\section{Background and Problem Statement}
In this section, the overall background of two popular admittance-modeling approaches, namely measurement-based and analytical methods, are presented in Section~\ref{Sec_IIP}. After that, the main challenges of multi-site admittance modeling with limited data are discussed in Section~\ref{Sec_Challenges}.
\subsection{Admittance} 
\begin{figure}[htbp]
\centerline{\includegraphics[scale = 0.85]{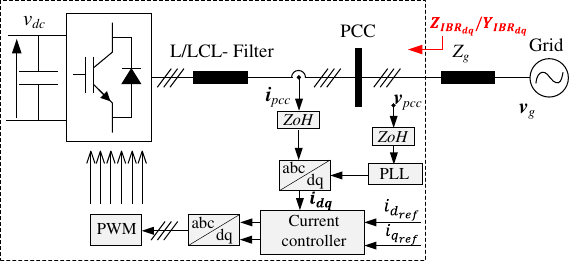}}
\caption{Studied grid-following IBR configuration. The combined hardware (e.g., L/LCL filter) and control loops (e.g., PLL, current controller, PWM) are treated as a black-box dq-admittance seen at the PCC for small-signal analysis.}
\label{fig_GFLIConfig}
\end{figure}

Identification Preliminaries\label{Sec_IIP}
Fig.~\ref{fig_GFLIConfig} shows a typical grid-following inverter (GFLI) configuration, including key components such as PLL, current controller, Pulse Width Modulator (PWM), and an L/LCL filter, which is the IBR configuration studied in this paper. The combined input--output behavior of these components at a given operating point can be represented by an impedance or admittance model. Such a model captures the small-signal dynamics, i.e., the current/voltage response to small perturbations around the operating point. Two common approaches to obtain IBR admittance models are introduced in the following subsections.

\subsubsection{Analytical admittance} model
Analytical admittance models of IBRs have been extensively studied \cite{wen2015analysis, zhou2022comprehensive, xu2023small, harnefors2007input}. Among the available techniques, the component-connection method is widely adopted due to its systematic formulation and practical implementation \cite{harnefors2007input, zhou2022comprehensive, xu2023small}. In this method, linearized subsystem blocks, each described in state-space form, are combined into an overall system by stacking the state, input, and output vectors and assembling the corresponding matrices in a block-diagonal form. Interconnection relationships among subsystem inputs and outputs are then imposed, from which the composite system matrices $\boldsymbol{A_{\text{sys}}}$, $\boldsymbol{B_{\text{sys}}}$, $\boldsymbol{C_{\text{sys}}}$, and $\boldsymbol{D_{\text{sys}}}$ are obtained using standard matrix algebra \cite{xu2023small}.

Once the overall state-space model is available, the transfer function from the system inputs (including $\boldsymbol{v_{dq}}$) to the outputs (including $\boldsymbol{i_{dq}}$) follows directly from control theory:

\begin{equation}
    \boldsymbol{G}(s) = \boldsymbol{C_{\text{sys}}}\big(s\boldsymbol{I_{\text{sys}}} - \boldsymbol{A_{\text{sys}}}\big)^{-1}\boldsymbol{B_{\text{sys}}} + \boldsymbol{D_{\text{sys}}},
\end{equation}
where $\boldsymbol{I_{\text{sys}}}$ is the identity matrix with the same dimension as $\boldsymbol{A_{\text{sys}}}$. The dq-admittance model is then obtained as the relevant $2\times2$ submatrix mapping $\boldsymbol{v_{dq}}$ to $\boldsymbol{i_{dq}}$, and can be expressed as
\begin{equation}
    \mathbf{Y}_{dq}(s) = \frac{\mathbf{i}_{dq}(s)}{\mathbf{v}_{dq}(s)}.
\end{equation}

Nevertheless, analytical modeling requires detailed information about the IBR hardware and software. In practice, such information is often difficult to obtain because limited standardization and proprietary considerations lead vendors to treat controller designs and parameters as commercial secrets.
\subsubsection{Admittance} measurement
To address the confidentiality issue, admittance models of black-box IBRs can be obtained using measurement-based approaches, either from physical hardware experiments or from EMT simulation models. A widely used method for three-phase IBRs in the dq domain is described in \cite{francis2011algorithm}. In this approach, a current perturbation source is connected in parallel at the point of common coupling (PCC). Before injecting perturbations, the system is driven to the desired steady-state operating point. Then, two independent small-signal current perturbations at a selected frequency $f_i$ are injected in the dq domain. The perturbation magnitude is chosen to be sufficiently small to avoid exciting nonlinear behavior, while remaining large enough to achieve an acceptable signal-to-noise ratio.

The instantaneous PCC voltage and current are measured and transformed into the dq frame as $v_{pcc_d}^{(1,2)}$, $v_{pcc_q}^{(1,2)}$, $i_{pcc_d}^{(1,2)}$, and $i_{pcc_q}^{(1,2)}$, where the superscripts $(1)$ and $(2)$ denote the two injection cases. The corresponding phasors at $f_i$ are extracted using the FFT as $V_{pcc_d}^{(1,2)}(f_i)$, $V_{pcc_q}^{(1,2)}(f_i)$, $I_{pcc_d}^{(1,2)}(f_i)$, and $I_{pcc_q}^{(1,2)}(f_i)$. The dq-impedance/admittance matrix at frequency $f_i$ is then computed as
\begin{equation}
\mathbf{Z_{dq}(f_i)}
=
\mathbf{I_{dq}(f_i)}^{-1}
\mathbf{V_{dq}(f_i)},
\end{equation}
\begin{equation}
\mathbf{Y_{dq}(f_i)}
=
\mathbf{V_{dq}(f_i)}^{-1}
\mathbf{I_{dq}(f_i)},
\end{equation}
where 
\[
\mathbf{Y_{dq}(f_i)} = 
\begin{bmatrix}
Y_{dd}(f_i) & Y_{dq}(f_i)\\
Y_{qd}(f_i) & Y_{qq}(f_i)
\end{bmatrix}
\]
\[
\mathbf{I_{dq}(f_i)} =
\begin{bmatrix}
I_d(f_i)^{(1)} & I_q(f_i)^{(1)}\\
I_d(f_i)^{(2)} & I_q(f_i)^{(2)}
\end{bmatrix}
\]
\[
\mathbf{V_{dq}(f_i)} =
\begin{bmatrix}
V_d(f_i)^{(1)} & V_d(f_i)^{(2)}\\
V_q(f_i)^{(1)} & V_q(f_i)^{(2)}
\end{bmatrix}.
\]

Repeating this procedure over a set of frequencies yields the impedance model across the frequency range of interest, which can then be used for small-signal stability analysis.

\subsection{Limitations of existing methods} \label{Sec_Challenges}
\begin{figure}[htbp]
\centerline{\includegraphics[scale = 1]{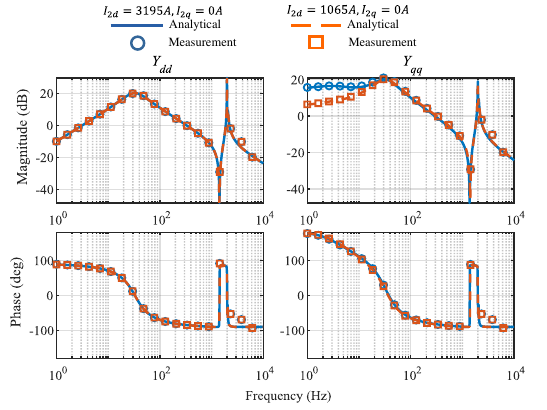}}
\caption{Analytical and measurement-based admittance magnitudes of the GFLI under different operating points.}
\label{fig_Admittance}
\vspace{-1.0\baselineskip}
\end{figure}

Although measurement-based admittance identification avoids the confidentiality limitations of analytical models, it introduces a major drawback in terms of time and computational cost. In particular, when investigating subsynchronous oscillations, the frequency range of interest for admittance characterization is typically 1--200~Hz. Consequently, EMT simulations or experiments must run for long durations to capture low-frequency dynamics, which makes the overall identification procedure expensive. Furthermore, IBR admittance characteristics depend nonlinearly on the operating point, as illustrated in Fig.~\ref{fig_Admittance}. Changes in dq currents (determined by the active and reactive power references) lead to variations in the admittance entries (e.g., $Y_{qq}$). Therefore, the measurement process must be repeated when analyzing different operating conditions.

Machine-learning-based methods have been studied to reduce the time and effort required for admittance identification. The core idea is that, given measurements at sparse frequency points and operating conditions, a trained model can predict admittance responses over a denser frequency grid and a broader set of operating points. However, existing approaches still require medium-to-large datasets from a single IBR for training, and most of them do not leverage multi-site data, as summarized in Table~\ref{Tab_Compar}. When the local dataset is extremely small, these methods are prone to overfitting.

\begin{table*}[!t]
\centering
\caption{Comparison of ML-based methods for admittance identification}
\label{Tab_Compar}
\setlength{\tabcolsep}{6pt}
\renewcommand{\arraystretch}{1.2}
\begin{tabularx}{\textwidth}{|p{4.5cm}|C{2.5cm}|Y|C{3.5cm}|C{1.5cm}|C{2cm}|}
\hline \hline
 Method & Data need (samples/IBR) & Generalization & Raw-data locality & Deployment (multi-site) & Number of IBRs \\ \hline
 Artificial Neural Network \cite{zhang2020artificial} & High (4800) & Medium & Low & Weak & Single\\ \hline
 Support Vector Machines \cite{mohammed2024support} & Medium (540) & Medium & Low & Weak & Single\\ \hline
 Recurrent Neural Network \cite{xiao2009recurrent} & High ($\approx 1000$) & Medium & Low & Weak & Single\\ \hline
 Physics-Informed NN \cite{zhang2022physics} & Medium-Low (200) & High & Low & Medium & Single\\ \hline
 Transfer Learning \cite{rane2025transfer, zhang2023transfer, li2024machine} & Medium-Low (300) & Medium--High & Medium/Low (centralized pretraining data) & Medium & Single \\ \hline
 Centralized ML & Low & High & Very low & Medium & Multiple \\ \hline
 \textbf{Proposed Federated Learning} & \textbf{Low (min 40)} & \textbf{High} & \textbf{High} & \textbf{High} & \textbf{Multiple} \\ 
 \hline \hline
\end{tabularx}
\end{table*}

A straightforward approach is to increase the training data by pooling measurements from multiple IBRs with similar characteristics. However, existing identification pipelines often use identical input grids across different IBRs (e.g., $\boldsymbol{V_1=V_2}$, $\boldsymbol{P_1=P_2}$, $\boldsymbol{Q_1=Q_2}$, and $\boldsymbol{f_1=f_2}$), which results in the same inputs but different outputs (i.e., different admittance profiles). In such a setting, simply aggregating more data does not necessarily improve learning effectiveness, because the model may struggle to explain conflicting input--output mappings without additional context that distinguishes each IBR.

Centralized learning can partially address data scarcity if the input sets are designed differently across IBRs (e.g., using different frequency grids, such as a linear spacing for $\boldsymbol{f_1}$ and a logarithmic spacing for $\boldsymbol{f_2}$). In that case, a centralized model can be trained on a more diverse dataset collected from multiple IBRs with the with different architectures and parameters, thereby reducing overfitting. Nevertheless, aggregating raw data in a central location raises privacy and cybersecurity concerns.

To address these issues, this paper proposes a federated-learning-based admittance identification framework tailored to extremely limited local datasets and strict data-locality requirements in systems with high IBR penetration.

\section{FedAvgIm: Federated Learning-Based Admittance Identification Framework}
\begin{figure*}[]
\centerline{\includegraphics[scale = 0.88]{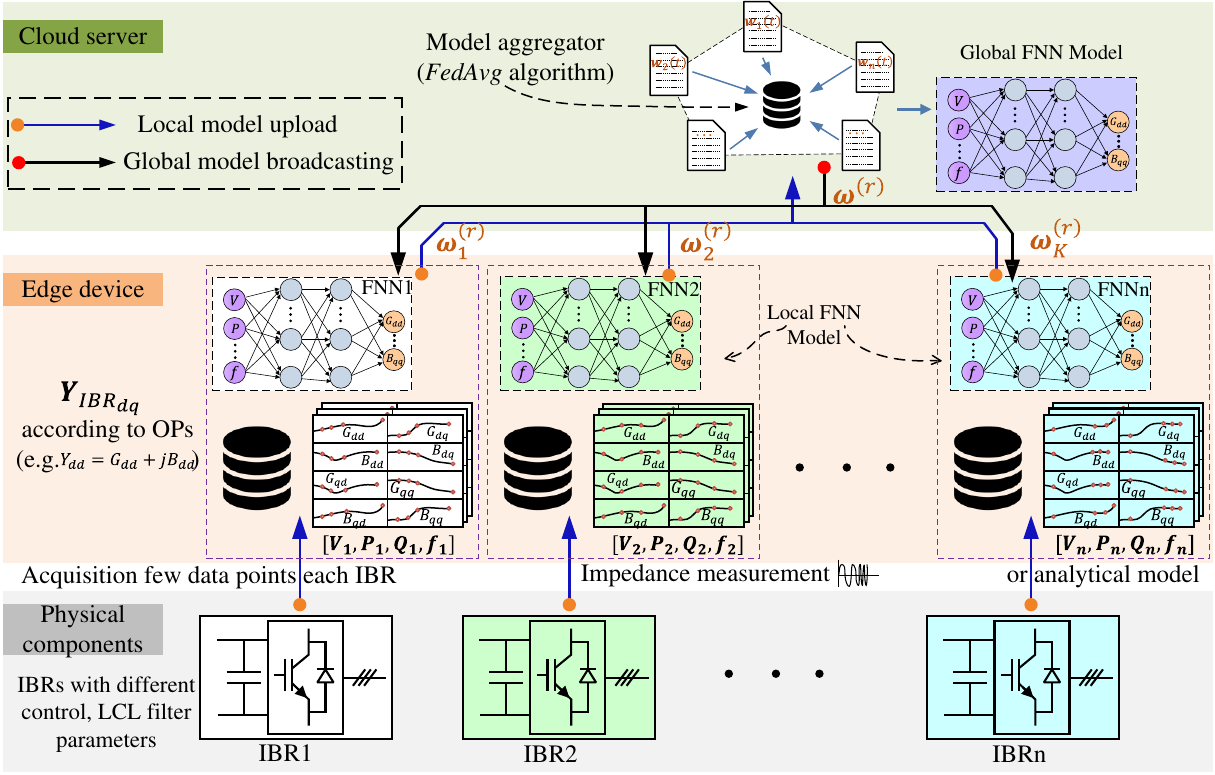}}
\caption{FedAvgIm framework for multi-site dq-admittance identification. Each IBR site acquires a small local dataset (from admittance measurement, EMT simulation, or analytical modeling) and trains a local FNN; only model parameters $w_k(t)$ are uploaded to the cloud aggregator, which computes the global model $w(t)$ and broadcasts it back to all clients, preserving data-locality while enabling cross-site learning.}
\label{fig_framework}
\end{figure*}

In this section, the proposed framework (Fig.~\ref{fig_framework}) is described in detail. Section~\ref{Sec_DA} presents the data collection and organization procedure. Next, the federated-learning-based identification algorithm is introduced, including the model architecture and key hyperparameters. Although dq-admittance is used as the learning target in this work, the corresponding impedance can be obtained when needed.

\subsection{Data Acquisition and Structure}\label{Sec_DA}
As shown at the bottom of Fig.~\ref{fig_framework}, admittance profiles can be obtained from EMT simulations, field devices using measurement-based methods \cite{francis2011algorithm}, or analytical IBR models \cite{li2024machine}. In this paper, an EMT simulation-based measurement approach is adopted to validate the proposed machine-learning framework under different measurement-noise levels and frequency-scan uncertainty conditions. The accuracy of the measurement-based admittance data is evaluated in Fig.~\ref{fig_Admittance} by comparison with the analytical admittance model.

In this work, two small linearly independent voltage perturbations are injected at the PCC in the dq frame, each with an amplitude of 5\% of the nominal voltage magnitude. A multitone frequency injection is used to reduce the measurement time. The instantaneous three-phase PCC voltages and currents are then measured and transformed into the dq frame. The dq admittance is estimated using the \textit{frestimate} function in MATLAB. This procedure is applied to nine grid-following inverters, denoted as IBR1--IBR9, with different topologies, such as L- and LCL-filter inverters, and different parameter settings. The detailed configurations and EMT simulation files are provided in \cite{Manh2025Appendix}. The dataset design follows domain consistency: the test set is made more challenging by using denser operating points and the full frequency set, while remaining within the same physical domain, i.e., $\text{ops}_{\text{train}} \subset \text{ops}_{\text{test}}$ and $f_{\text{train}} \subset f_{\text{test}}$.

Each input sample is defined as $(V, P, Q, f)$, where $V$, $P$, and $Q$ are per-unit operating-point variables and $f$ is the perturbation frequency in Hz. Importantly, the operating-point grids are not identical across IBRs. In particular, each IBR uses its own operating-point domain $(V,P,Q)$, defined by different ranges or customized operating-point patterns, before the frequency sweep is performed. Infeasible operating points are removed by enforcing the rated apparent-power constraint
\begin{equation}
    P^2 + Q^2 \le 1 \quad (\text{p.u.}).
\end{equation}
The test operating points are constructed by densifying the training domain, and the same power constraint is applied.

For each remaining operating point, the measurement procedure is performed using the predefined voltage magnitude $V$ and the corresponding steady-state current setpoints derived from $(P,Q)$. The dq-admittance matrix $Y(j\omega)\in\mathbb{C}^{2\times 2}$ is then estimated from EMT Simulink simulations over a prescribed frequency set $f_{\mathrm{test}}$, whose distribution may be logarithmic, piecewise-dense, or randomized depending on the IBR scenario. The training frequency set is obtained by fixed-stride subsampling of this predefined frequency grid. In the implementation, the subsampling is applied to the frequency-grid index, e.g., $f_{\mathrm{train}}=\{f_{\mathrm{test},i}\mid i=1,3,5,\ldots,N_f\}$, rather than to a newly generated uniformly linear frequency grid. Therefore, when $f_{\mathrm{test}}$ is logarithmic or denser in the low-frequency region, $f_{\mathrm{train}}$ preserves the corresponding low-frequency coverage. The released datasets and data-generation scripts in \cite{Manh2025Appendix} include the exact frequency vectors, selected training indices, and train/test split metadata for reproducibility.

The estimated dq-admittance matrix is used as the learning target. For each operating-point--frequency pair, the four dq entries are used as labels by stacking their real and imaginary parts:
\[
\mathbf{y}=
\begin{bmatrix}
\Re(\operatorname{vec}(Y_{dq}))\\
\Im(\operatorname{vec}(Y_{dq}))
\end{bmatrix}
\in\mathbb{R}^{8},
\]
where $\operatorname{vec}(\cdot)$ stacks the entries of the $2\times2$ matrix into a $4\times1$ vector.

Therefore, each dataset sample corresponds to one operating-point--frequency pair, with $x\in\mathbb{R}^{4}$ and $\mathbf{y}\in\mathbb{R}^{8}$. For a given dataset, the total number of samples equals $N_{\text{ops}}\times N_f$, where $N_{\text{ops}}$ is the number of feasible operating points retained after filtering and $N_f$ is the number of frequency points used, which is subsampled for training and fully used for testing.

Since the proposed framework targets data-scarce settings, each of the nine IBR datasets contains only a small number of samples. Specifically, the numbers of training samples are: IBR1 (390), IBR2/IBR3/IBR5 (330 each), IBR6 (220), IBR8 (180), IBR7 (140), IBR4 (100), and IBR9 (40). The total estimation time for the nine IBRs is approximately 7 hours on a laptop with a 13th Gen Intel(R) Core(TM) i5-13500HX CPU at 2.50 GHz and 16.0 GB RAM. For the full validation study, which requires admittance models under four different noise levels and two different IBR structures, the total estimation time is approximately 42 hours. These datasets are provided in \cite{Manh2025Appendix}. In the federated-learning setting considered here, the datasets are stored locally at the corresponding IBR sites, and their use for decentralized training is discussed in the next subsection.

\subsection{FedAvgIm: Federated Averaging for Multi-Site Admittance Identification}\label{Sec_FLA}
To reduce raw-data exposure, datasets are stored locally at each site (second layer of Fig.~\ref{fig_framework}) and are typically small. In this setting, decentralized learning methods such as federated learning are well suited. It should be noted that federated learning does not inherently guarantee privacy. In the proposed FedAvgIm framework, the privacy benefit mainly arises from data locality: raw measurement data remain on local clients and are not directly shared with the central coordinator, while only model parameters are exchanged. In this work, an honest-but-curious threat model is considered, in which the cloud aggregator and clients are assumed to follow the FedAvg protocol but may inspect the exchanged model parameters. Under this setting, FedAvgIm improves data locality by keeping proprietary admittance measurements at local IBR sites and avoiding centralized raw-data pooling. However, it does not provide formal privacy guarantees against inference attacks, such as model inversion, gradient leakage, or membership inference, nor against malicious behaviors such as poisoning or collusion.

Based on this data-locality setting, a federation of $K$ inverter sites is considered, where client $k$ holds a private dataset
$\mathcal{D}_k=\{(x_i,y_i)\}_{i=1}^{n_k}$ generated from local operating-point selections and frequency sweeps.
Each input sample is defined as
$x=(V,P,Q,f)\in\mathbb{R}^{4}$, where $V$, $P$, and $Q$ are per-unit operating-point variables and $f$ is the perturbation frequency.
The learning target is the dq-admittance matrix $Y(j\omega)\in\mathbb{C}^{2\times 2}$.
Following common practice in impedance/admittance learning, the complex-valued dq entries are represented by stacking their real and imaginary parts into an $8$-dimensional real vector $\mathbf{y}\in\mathbb{R}^{8}$. Therefore, each operating-point--frequency pair has input $x\in\mathbb{R}^{4}$ and label $\mathbf{y}\in\mathbb{R}^{8}$.

\paragraph*{Model and local objective}
At each local site, an FNN predictor $f(\cdot;w)$ is trained to map the operating-point--frequency input
$x\in\mathbb{R}^{4}$ to the admittance label $\mathbf{y}\in\mathbb{R}^{8}$.
The model parameters are denoted by $w$, and all parameters are shared and updated in the federated training process.
The FNN used in this work consists of fully connected layers with nonlinear activation functions:
\begin{equation}
\hat{\mathbf{y}} = f(x;w).
\end{equation}

At local site $k$, the client minimizes the mean-squared error (MSE) over its private dataset
$\mathcal{D}_k$:
\begin{equation}
F_k(w)=\frac{1}{n_k}\sum_{(x,\mathbf{y})\in\mathcal{D}_k}
\left\|f(x;w)-\mathbf{y}\right\|_2^2 .
\end{equation}
The global objective is defined as the sample-size-weighted average of the local objectives:
\begin{equation}
F(w)=\sum_{k=1}^{K}\frac{n_k}{n}F_k(w),
\qquad
n=\sum_{k=1}^{K}n_k .
\end{equation}

\begin{algorithm}[!b]
\caption{FedAvgIm: FedAvg for dq-admittance learning}
\label{alg:fedavgim}
\begin{algorithmic}[1]
\Statex \textbf{Input:} Local admittance datasets $\{\mathcal{D}_k\}_{k=1}^{K}$ with sizes $\{n_k\}_{k=1}^{K}$; communication rounds $R$; local epochs $E$; batch size $B$; client fraction $C$; learning-rate schedule $\{\eta_r\}_{r=1}^{R}$.
\Statex \textbf{Output:} Global model parameters $w^{(R)}$.

\State Initialize global model parameters $w^{(0)}$.
\For{$r = 1$ to $R$}
    \State Select a subset of clients $\mathcal{S}_r \subseteq \{1,\dots,K\}$ according to client fraction $C$.
    \State Broadcast $w^{(r-1)}$ to all selected clients $k\in\mathcal{S}_r$.
    \ForAll{$k \in \mathcal{S}_r$ \textbf{in parallel}}
        \State Initialize local parameters $w_k^{(r,0)} \leftarrow w^{(r-1)}$.
        \State Set $w_k \leftarrow w_k^{(r,0)}$.
        \For{$e = 1$ to $E$}
            \ForAll{mini-batches $(x,\mathbf{y})$ from $\mathcal{D}_k$ of size $B$}
                \State Update all parameters $w_k$ using Adam to minimize
                $\left\|f(x;w_k)-\mathbf{y}\right\|_2^2$ with step size $\eta_r$.
            \EndFor
        \EndFor
        \State Send updated local parameters $w_k^{(r)} \leftarrow w_k$ to the server.
    \EndFor
    \State Aggregate the received models using sample-size-weighted averaging:
    \Statex \hspace{\algorithmicindent}
    $w^{(r)} \leftarrow
    \sum_{k\in\mathcal{S}_r}
    \frac{n_k}{\sum_{j\in\mathcal{S}_r} n_j}\, w_k^{(r)}$.
\EndFor
\State \textbf{return} $w^{(R)}$.
\end{algorithmic}
\end{algorithm}

\begin{table}[h]
\caption{Hyperparameters and model configuration for FedAvgIm}
\label{tab:hyperparams}
\centering
\renewcommand{\arraystretch}{1.05}
\setlength{\tabcolsep}{3pt}
\begin{tabular}{p{0.3\columnwidth} p{0.6\columnwidth}}
\hline \hline
\textbf{Item} & \textbf{Value} \\
\hline
\multicolumn{2}{l}{\emph{Model architecture}}\\
Input dimension  & $4$ \\
Output dimension  & $8$ \\
Hidden dimension  & $32$ \\
FNN architecture & Linear($4\!\rightarrow\!32$) + ReLU + Linear($32\!\rightarrow\!16$) + ReLU + Linear($16\!\rightarrow\!8$) \\
Loss function & MSE \\
Optimizer & Adam \\
\hline
\multicolumn{2}{l}{\emph{Federated training}}\\
Federated rounds  & $250$ \\
Local epochs per round  & $8$ \\
Local batch size  & $64$ \\
Evaluation batch size & $512$ \\
Learning rate & $10^{-3}$ \\
Random seed & $45$ \\
\hline \hline
\end{tabular}
\end{table}

\paragraph*{Server aggregation}

The cloud aggregator employs the standard FedAvg algorithm, a synchronous federated optimization procedure \cite{mcmahan2017communication}.
At communication round $t$, the server broadcasts the current global model $w_t$ to the selected clients.
Each selected client initializes its local model with $w_t$ and updates all model parameters using multiple local epochs of mini-batch training on its private dataset.
After local training, the clients send the updated model parameters back to the server.
The server then computes the next global model using a sample-size-weighted average:
\begin{equation}
w_{t+1}=\sum_{k\in\mathcal{S}_t}\frac{n_k}{\sum_{j\in\mathcal{S}_t}n_j}w_{t+1}^{k},
\end{equation}
where $\mathcal{S}_t$ is the set of selected clients at round $t$, and $w_{t+1}^{k}$ denotes the locally updated model parameters from client $k$.
The detailed procedure is summarized in Algorithm~\ref{alg:fedavgim}, and the model/training hyperparameters are listed in Table~\ref{tab:hyperparams}.

\paragraph*{Communication and computational overhead} 
The communication cost of FedAvgIm is determined by the size of the exchanged FNN parameters \cite{cui2021slashing}. For the FNN in Table~II, the number of trainable parameters is
\[
|w|=(4\times32+32)+(32\times16+16)+(16\times8+8)=824.
\]
Assuming 32-bit floating-point parameters, the model size is approximately 3.30~kB. In each communication round, each selected client downloads the global model and uploads the locally updated model. Therefore, under the full-participation setting used in this work with nine clients and 250 communication rounds, the total communication volume is approximately 7.42~MB uplink, 7.42~MB downlink, and 14.83~MB in total. Compared with local-only learning, FedAvgIm introduces additional model-exchange communication, but enables cross-site learning without raw-data pooling. Compared with centralized learning, FedAvgIm avoids transferring proprietary raw admittance datasets to a central server, at the cost of repeated lightweight model-parameter exchange.

The computational cost is also modest because the model contains only 824 trainable parameters and each client has a small local dataset. With 250 communication rounds, 8 local epochs per round, and batch size 64, the largest client with 390 samples performs approximately 14,000 local mini-batch updates, while the smallest client with 40 samples performs approximately 2,000 updates. The server-side computation is limited to sample-size-weighted averaging of the received model parameters. Therefore, the main computational burden of the overall study comes from EMT-based admittance data generation rather than from training the lightweight FNN.

\section{Performance Evaluation}
Several evaluation cases are considered to assess the proposed framework under extremely data-scarce local IBR settings, noisy measurement environments, and non-IID scenarios:
\begin{itemize}
    \item Case 1: Comparison with fully centralized training and local-only training across nine clients.
    \item Case 2: Impact of adding a new IBR client during training.
    \item Case 3: Comparison with transfer learning under strict data-locality constraints.
    \item Case 4: Evaluation under measurement-noise conditions.
    \item Case 5: Efficiency evaluation under different non-IID levels.
    \item Case 6: Application of the estimated admittance models to IBR--grid small-signal stability analysis.
\end{itemize}
Specifically, the datasets used in Cases 1--4 are obtained from nine IBRs with the same topology but different parameter settings. In Case 5, some IBRs are modified from LCL-filter to L-filter configurations to introduce stronger physical heterogeneity and evaluate the framework under more pronounced non-IID conditions.

\subsection{Case 1: Comparison with fully centralized training and local-only training across nine clients}

\begin{figure}[!h]
\centerline{\includegraphics[scale = 1]{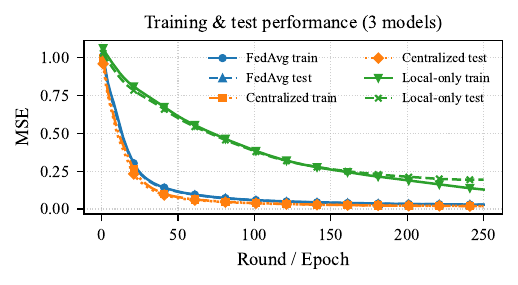}}
\caption{Training and test MSE learning curves over 250 rounds/epochs for the three approaches (Centralized, FedAvgIm, and Local-only). FedAvgIm converges close to the centralized baseline while reducing the train--test gap relative to Local-only training.}
\label{fig_Case1_learningcurve}
\end{figure}

In this scenario, the centralized, federated-learning, and local-only approaches share the same model architecture and hyperparameters, and are trained for 250 epochs. The training and test learning curves are shown in Fig.~\ref{fig_Case1_learningcurve}. The centralized model converges faster, reaching a plateau after approximately 100-150 epochs, since it is trained on aggregated data and therefore benefits from richer supervision and improved generalization. After 200 rounds, the proposed federated-learning model achieves performance close to the centralized baseline in terms of MSE. By contrast, the local-only curve (reported as the average MSE across the nine IBRs at each epoch) shows early signs of overfitting, as the training and test curves gradually diverge.

\begin{figure}[h]
\centerline{\includegraphics[scale = 1]{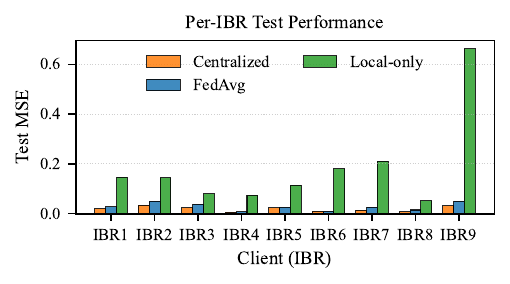}}
\caption{Per-IBR test MSE comparison for the three training strategies (Centralized, FedAvgIm, and Local-only). FedAvgIm achieves errors close to centralized training and consistently outperforms Local-only across all nine IBRs, with the largest gains observed for the most data-scarce clients (e.g., IBR9).}
\label{fig_Case1_bar}
\end{figure}

The final test MSE for each IBR is reported in Fig.~\ref{fig_Case1_bar}. The centralized model achieves the lowest MSE across all nine IBRs, followed by the proposed federated-learning model, whose MSE values are close to the centralized baseline. Both approaches outperform the local-only setting, where each IBR trains its own FNN using only local data, with the largest performance gap observed for IBR9, which has only 40 training samples.

\begin{figure}[h]
\centerline{\includegraphics[scale = 1]{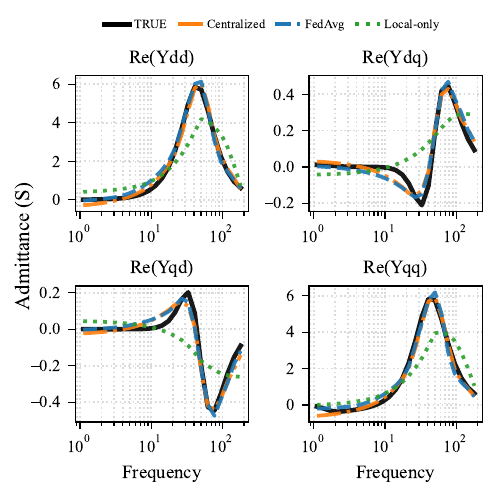}}
\caption{Estimated versus ground-truth dq-admittance frequency responses for real part of IBR1 of four entries ($Y_{dd}$, $Y_{dq}$, $Y_{qd}$, $Y_{qq}$) over frequency, comparing Centralized, FedAvgIm, and Local-only models against the true response.}
\label{fig_Case1_Impedance}
\end{figure}

For illustration, Fig.~\ref{fig_Case1_Impedance} compares the estimated and ground-truth real parts of IBR1 admittance profiles at $V=1$~p.u., $P=1$~p.u., and $Q=0$ on the test set. The federated-learning model achieves accuracy close to the centralized baseline and tracks the ground truth well across frequency. By contrast, the local-only model exhibits larger errors due to limited training data.

\begin{figure}[h]
\centerline{\includegraphics[scale = 1]{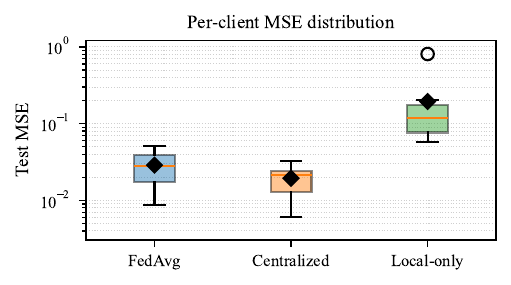}}
\caption{Boxplot of per-client test MSE (log scale) comparing Centralized, FedAvg, and Local-only models. Centralized and FedAvgIm achieve lower and more consistent errors across clients, whereas Local-only exhibits higher variability and outliers.}
\label{fig_Case1_box}
\end{figure}

The results in Fig.~\ref{fig_Case1_bar} are further summarized in Fig.~\ref{fig_Case1_box}, which shows the distribution of per-client test MSE (log scale) for the three approaches. The centralized model attains the lowest median MSE with a narrow interquartile range, indicating both strong performance and high consistency across clients. FedAvgIm achieves results close to the centralized approach, with only a slight increase in the median while maintaining low variability. In contrast, the local-only approach exhibits a substantially higher median MSE and much wider dispersion with outliers, reflecting unstable training and poor generalization when each client is trained solely on limited local data.

\begin{figure}[h]
\centerline{\includegraphics[scale = 1]{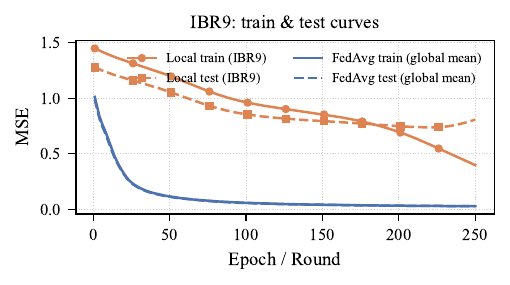}}
\caption{Training and test MSE curves for client IBR9, comparing the Local-only model with the global mean performance of FedAvgIm. FedAvgIm converges faster and achieves lower test error, while the Local-only model exhibits higher error and signs of overfitting.}
\label{fig_Case1_IBR9}
\end{figure}

To examine the extreme data-scarce case, the learning curves of the local-only FNN for IBR9 are compared with the FedAvgIm performance in Fig.~\ref{fig_Case1_IBR9}. With only 40 samples, the local-only model suffers from overfitting: the training error continues to decrease while the test error starts to increase after approximately epoch 200. By contrast, federated learning mitigates overfitting by leveraging shared information across clients, leading to faster convergence and lower test error.

\subsection{Case 2: Impact of adding a new IBR client during training}\label{Sec_NewIBR}
\begin{figure}[h]
\centerline{\includegraphics[scale = 1]{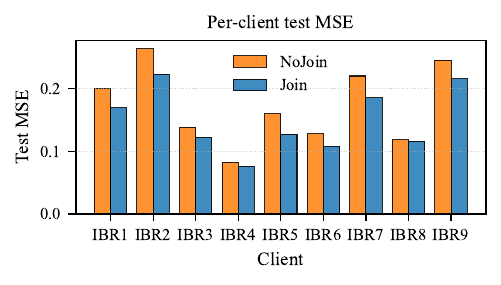}}
\caption{Per-client test MSE comparison under the NoJoin and Join settings. The Join strategy achieves lower test errors across all clients (IBR1--IBR9), indicating improved generalization compared to the NoJoin case.}
\label{fig_Case2_bar}
\end{figure}

\begin{figure}[h]
\centerline{\includegraphics[scale = 1]{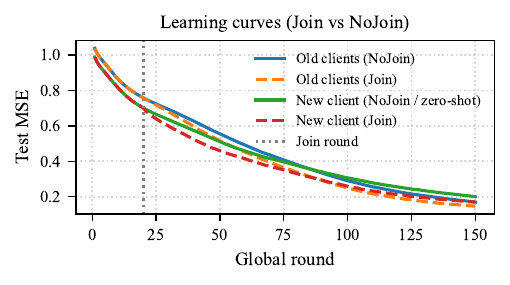}}
\caption{ Test MSE curves for the scenario in which a new IBR joins the federation. The curves show test MSE over global rounds for existing and joining clients; the dashed vertical line indicates the join round, after which the Join setting exhibits improved convergence and reduced error compared to the NoJoin setting.}
\label{fig_Case2_LearningCurve}
\end{figure}

\begin{figure}[h]
\centerline{\includegraphics[scale = 1]{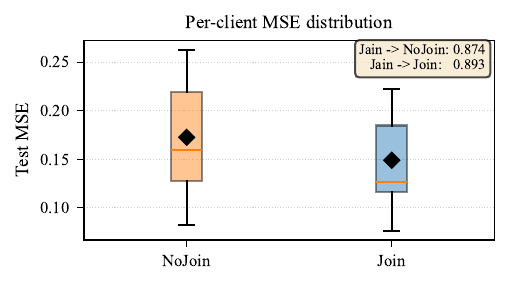}}
\caption{Boxplot of per-client test MSE comparing the NoJoin and Join settings. Enabling client joining reduces both the median error and the variability across clients, indicating improved overall performance and stability.}
\label{fig_Case2_boxplot}
\end{figure}

To investigate learning behavior when a new IBR joins the federation, the proposed federated-learning framework is first trained with eight clients (IBR2--IBR9), excluding IBR1. At round 20, a new client (IBR1) joins the training process. The results are shown in Fig.~\ref{fig_Case2_bar}, Fig.~\ref{fig_Case2_LearningCurve}, and Fig.~\ref{fig_Case2_boxplot}.

Fig.~\ref{fig_Case2_bar} summarizes the impact of client joining by reporting the per-client test MSE under the Join and NoJoin settings. The Join setting yields lower test errors for every IBR, indicating that integrating a new IBR with the same structure consistently improves performance. The learning dynamics are further illustrated in Fig.~\ref{fig_Case2_LearningCurve}, which compares the test MSE trajectories for the existing clients (IBR2--IBR9) and the joining client (IBR1) over 150 rounds. After the join event (round 20), both the existing clients and the joining client converge faster than in the NoJoin setting. Since the joining IBR shares the same characteristics as the existing clients, adding it effectively increases the available training data, benefiting all clients through improved generalization.

In addition, Fig.~\ref{fig_Case2_boxplot} shows that the Join setting reduces not only the mean error but also the dispersion across clients as the interquartile range is slightly narrower than that of the NoJoin case, indicating improved stability. To quantify fairness across clients, Jain's fairness index is computed from the per-client test MSE values; $J\in(0,1]$, and larger values indicate more uniform performance across clients.

\subsection{Case 3: Comparison with transfer learning under strict data-locality}  constraints\label{Sec_Transfer}

Transfer learning is a common approach for impedance/admittance identification \cite{li2024machine, zhang2023transfer}, where a model pretrained on one IBR is reused as an initialization for a different target IBR, referred to here as local transfer learning. This subsection compares local transfer learning with federated-learning-based transfer learning (FL-TL) under data-locality constraints and limited labeled data at the target IBR (IBR1). The two approaches use the same network architecture and fine-tuning hyperparameters. In this scenario, local transfer learning is pretrained using the IBR4 dataset for 150 epochs, whereas FL-TL is pretrained in a distributed manner using eight client datasets (IBR2--IBR9) for 150 communication rounds. Fine-tuning is performed for 64 epochs using 25\% of the IBR1 dataset, corresponding to 98 labeled samples.

It should be noted that the local TL and FL-TL settings differ in the amount and diversity of pretraining data. Therefore, this comparison is intended to evaluate a practical data-locality-preserving transfer-learning scenario rather than to isolate the effect of the FedAvgIm optimizer under equal pretraining data size. In local TL, the target model reuses a model pretrained from a single source IBR, whereas FL-TL allows multiple source IBRs to collaboratively pretrain a shared initialization without centralizing their raw admittance datasets. Thus, the improvement of FL-TL reflects the benefit of federated multi-source pretraining under raw-data-locality constraints, rather than the effect of FedAvgIm alone.

\begin{figure}[h]
\centerline{\includegraphics[scale = 1]{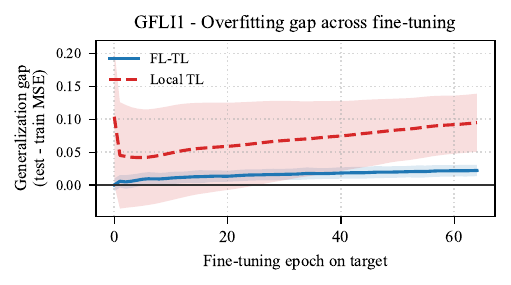}}
\caption{Overfitting gap during fine-tuning on GFLI1, measured as the difference between test and train MSE. FL-TL maintains a small and stable generalization gap, whereas local TL exhibits a larger and increasing gap, indicating stronger overfitting and weaker generalization under limited target data.}
\label{fig_Case3_OverfittingCurve}
\end{figure}

Fig.~\ref{fig_Case3_OverfittingCurve} compares the overfitting behavior of FL-TL and local transfer learning during fine-tuning on GFLI1 under severe data scarcity. The generalization gap, defined as the difference between test and train MSE, is used as an indicator of overfitting. FL-TL maintains a small and stable generalization gap throughout the fine-tuning process, suggesting better generalization to the target data. By contrast, local transfer learning exhibits a larger and gradually increasing gap after the initial epochs, indicating stronger overfitting and weaker generalization when only limited target data are available.

\begin{figure}[h]
\centerline{\includegraphics[scale = 1]{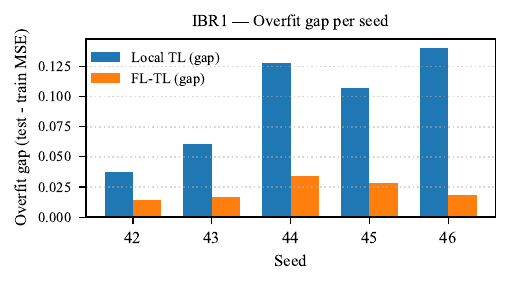}}
\caption{Overfitting gap (test--train MSE) on GFLI1 across different random seeds, comparing local transfer learning and FL-based transfer learning; FL-TL consistently yields a smaller overfitting gap, demonstrating more stable and robust generalization across runs.}
\label{fig_Case3_OverfittingBar}
\end{figure}

The overfitting gap, defined as the difference between the final test and training MSE, is shown in Fig.~\ref{fig_Case3_OverfittingBar} for five random seeds. Although the absolute values vary across seeds, FL-TL consistently exhibits a smaller overfitting gap than local transfer learning, indicating more stable and robust generalization across runs. It should be noted, however, that this comparison reflects a practical multi-source federated pretraining setting rather than a controlled experiment with matched pretraining data size. A data-size-controlled comparison would be useful to further separate the contribution of pretraining data diversity from that of federated optimization and will be considered in future work.

\subsection{Case 4: Evaluation under measurement-noise conditions}\label{Sec_Noise}

In this scenario, four levels of white measurement noise ($0\%$, $1\%$, $2\%$, and $3\%$) are applied to the voltage and current measurements before the admittance profiles are estimated. The total time required for the data collection process is approximately 28 hours.

\begin{figure}[h]
\centerline{\includegraphics[scale = 1]{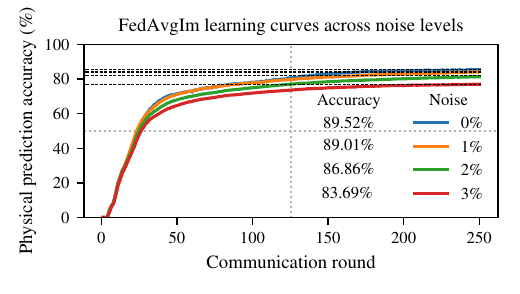}}
\caption{FedAvgIm learning curves under different measurement-noise levels. Although the final test accuracy decreases as the noise level increases, all cases exhibit stable convergence within approximately 150 communication rounds.}
\label{fig_Case4_Noise}
\end{figure}

As shown in Tab.~\ref{tab:noise_test} and Fig.~\ref{fig_Case4_Noise}, the proposed framework maintains stable learning behavior under all tested measurement-noise levels. In the noise-free case, FedAvgIm achieves a test MSE of $0.0408$ and a physical prediction accuracy of $89.52\%$. When the noise level increases to $1\%$, the accuracy remains nearly unchanged at $89.01\%$, indicating that the proposed framework is insensitive to mild measurement noise. As the noise level further increases to $2\%$ and $3\%$, the physical prediction accuracy gradually decreases to $86.86\%$ and $83.69\%$, respectively. This trend is expected because noise in the measured voltage and current signals affects the quality of the estimated admittance labels. Nevertheless, FedAvgIm still retains more than $83\%$ physical prediction accuracy even under the highest tested noise level, demonstrating its robustness in noisy measurement environments.

\begin{table}[h]
\caption{Performance under different measurement-noise levels}
\label{tab:noise_test}
\centering
\renewcommand{\arraystretch}{1.05}
\setlength{\tabcolsep}{6pt}
\begin{tabular}{c c c}
\hline \hline
\textbf{Noise levels} & \textbf{Test MSE} & \textbf{Physical prediction accuracy}\\
\hline
0\%  & $0.0408$  & $89.52\%$\\
1\%  & $0.0519$  & $89.01\%$\\
2\%  & $0.0862$  & $86.86\%$\\
3\%  & $0.1389$  & $83.69\%$\\
\hline \hline
\end{tabular}
\end{table}

Fig.~\ref{fig_Case4_Noise} further illustrates the convergence behavior of FedAvgIm under different measurement-noise levels using the physical prediction accuracy metric. The accuracy is computed after inverse-transforming the predicted admittance responses back to physical units and evaluating the normalized prediction error. Although higher noise levels lead to lower final accuracy, all learning curves converge smoothly within approximately 150 communication rounds. This indicates that measurement noise mainly affects the attainable prediction accuracy rather than destabilizing the federated training process. Overall, the results show that FedAvgIm provides stable and robust admittance identification performance under noisy measurement conditions.

\subsection{Case 5: Efficiency evaluation under different non-IID
levels}\label{Sec_nonIID}

Statistical heterogeneity, or non-IID data, is an important issue in federated learning frameworks, particularly for IBR admittance estimation, where IBRs may be provided by different vendors and may have different circuit structures. To investigate this issue, three cases (G1, G2, and G3) with different levels of non-IIDness are designed by varying the IBR topology and parameters. Specifically, G1 consists of nine IBRs with the same circuit structure but different parameters. G2 represents a mixed-topology case, in which IBR4 and IBR9 use L-filter inverters, while the remaining IBRs use LCL-filter inverters. G3 has the same topology configuration as G2 but introduces significantly larger parameter variations. The maximum mean discrepancy (MMD) metric is used to quantify the distributional discrepancy among the datasets and assess whether their distributions are similar \cite{gretton2012kernel}. In this study, G1 is used as the baseline case. A summary of the designed cases is shown in Table~\ref{tab:MMD}.

\begin{table}[h]
\caption{Description of the non-IID heterogeneity study}
\label{tab:MMD}
\centering
\renewcommand{\arraystretch}{1.15}
\setlength{\tabcolsep}{3pt}
\footnotesize
\begin{tabularx}{\columnwidth}{c X c}
\hline\hline
\textbf{Case} & \centering\arraybackslash\textbf{Description} & \textbf{MMD increase} \\
\hline
G1 & Same structure with different parameters & Baseline \\
G2 & Mixed structure; IBR4 and IBR9 use L-filter inverters & $1.17\%$ \\
G3 & Same structure as G2 with significantly larger parameter variations & $5.74\%$ \\
\hline\hline
\end{tabularx}
\end{table}

\begin{figure}[h]
\centerline{\includegraphics[scale = 1]{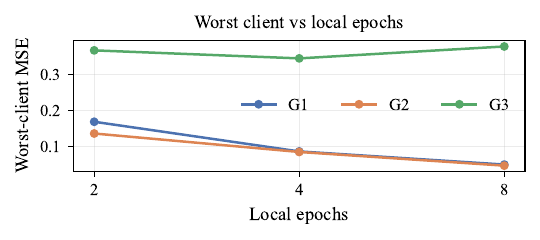}}
\caption{Worst-client test MSE versus local epochs under different heterogeneity levels. Increasing local epochs improves the worst-client performance in G1 and G2, whereas G3 remains substantially worse due to severe client heterogeneity and client drift.}
\label{fig:Case5_worst}
\end{figure}

Fig.~\ref{fig:Case5_worst} evaluates the worst-client test MSE under different numbers of local epochs. For G1 and G2, increasing the number of local epochs reduces the worst-client MSE, indicating that additional local computation helps improve the performance of the most poorly performing client. This suggests that, under mild or moderate heterogeneity, increasing local training can enhance the fairness of FedAvgIm by allowing each client to better adapt to its local admittance characteristics before aggregation. However, this trend is not observed in G3. The worst-client MSE in G3 remains much higher than those in G1 and G2 for all local epoch settings, and increasing the local epoch does not lead to a consistent improvement. Since G3 has the largest parameter variations and the highest MMD increase, this result indicates that severe statistical heterogeneity can cause stronger client drift, where local models move toward different client-specific optima before aggregation. Consequently, the aggregated global model becomes less effective for the worst-performing client.

\begin{figure}[h]
\centerline{\includegraphics[scale = 1]{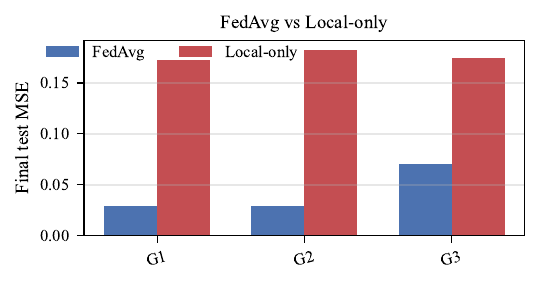}}
\caption{Comparison of final test MSE between FedAvgIm and local-only training under different heterogeneity levels. FedAvgIm consistently outperforms local-only training, although its advantage is reduced in the highly heterogeneous G3 case.}
\label{fig:Case5_AvgMSE}
\end{figure}

Fig.~\ref{fig:Case5_AvgMSE} compares the final test MSE of FedAvgIm and local-only training under the three heterogeneity cases. FedAvgIm consistently achieves a lower test MSE than local-only training, demonstrating the benefit of collaborative learning across IBR datasets. In G1 and G2, the improvement is substantial because the clients share relatively similar admittance characteristics, allowing the global model to learn common patterns across clients. In contrast, although FedAvgIm still outperforms local-only training in G3, the performance gap is reduced. This agrees with the MMD-based heterogeneity analysis, which shows that G3 has the largest distributional discrepancy among clients. Therefore, the results indicate that FedAvgIm is effective for improving generalization compared with isolated local training, but its advantage decreases when client heterogeneity becomes severe.

\begin{figure}[h]
\centering
\subfloat[G1: low-heterogeneity scenario.]{
    \includegraphics[width=0.95\linewidth]{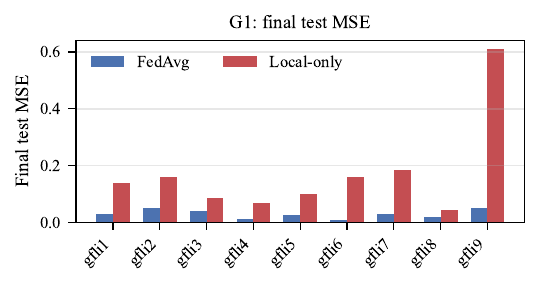}
    \label{fig:G1_client_final_test_mse}
}
\vspace{1mm}

\subfloat[G3: highly heterogeneous scenario.]{
    \includegraphics[width=0.95\linewidth]{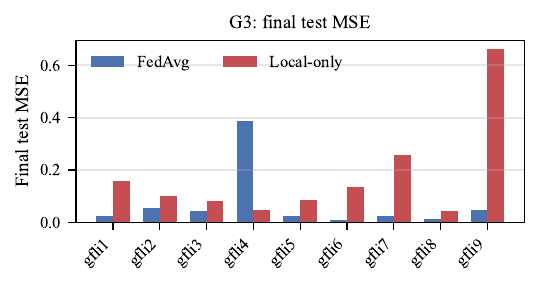}
    \label{fig:G3_client_final_test_mse}
}
\caption{Per-client final test MSE comparison between FedAvgIm and local-only training in G1 and G3. FedAvgIm achieves consistently low errors across clients in the low-heterogeneity G1 case. In the highly heterogeneous G3 case, FedAvgIm still improves most clients compared with local-only training, but the large error of gfli4 indicates client-level unfairness caused by severe heterogeneity and client drift.}
\label{fig:client_final_test_mse_fedavg_vs_local}
\end{figure}

Fig.~\ref{fig:client_final_test_mse_fedavg_vs_local} compares the per-client final test MSE of FedAvgIm and local-only training in G1 and G3. In the low-heterogeneity G1 case, FedAvgIm achieves consistently low errors across nearly all clients and clearly outperforms local-only training. This demonstrates that, when IBRs share the same circuit structure and only differ in parameters, the global model learned by FedAvgIm can effectively capture common admittance characteristics across clients. By contrast, local-only training produces larger and more uneven errors, especially for gfli9, indicating that isolated training is more sensitive to limited local data and client-specific variations.

In the highly heterogeneous G3 case, FedAvgIm still provides performance gains for most clients compared with local-only training, but its benefit becomes less uniform. In particular, gfli4 shows a much higher FedAvgIm test MSE than the other clients. This result suggests that strong topology and parameter heterogeneity can make some client distributions deviate significantly from the global distribution learned during aggregation. Consequently, the aggregated model may be biased toward the dominant client patterns and may not generalize well to outlier clients. This supports the observation from the worst-client analysis and the MMD-based heterogeneity study that severe non-IIDness can introduce client drift and reduce the fairness of FedAvgIm across clients.

It should be noted that a single surrogate model is unlikely to generalize across all types of inverter-based resources (IBRs), given their diverse control architectures and converter topologies. Instead, the proposed framework is intended to operate effectively across groups of IBRs with similar behavioral characteristics. To systematically address broader heterogeneity, future work should consider multi-model frameworks based on clustered federated learning or personalized federated learning \cite{tan2022towards}.

\subsection{Case 6: Application of the estimated admittance models to IBR--grid small-signal stability analysis}\label{Sec_Stability}

A black-box IBR can be represented by an impedance/admittance model, which captures its small-signal dynamics at a given operating point, i.e., the current/voltage response to small perturbations. A common application of such models is small-signal stability analysis of IBR--grid interactions. In this framework, the IBR is modeled as an equivalent admittance, together with an appropriate controlled-source representation, interfaced with the grid impedance and an ideal voltage source \cite{sun2011impedance}. By defining the return-ratio matrix $\boldsymbol{L}(s)$ in \eqref{eq_return}, the two eigenvalues $\lambda_1(s)$ and $\lambda_2(s)$ can be computed and used to construct generalized Nyquist plots for stability assessment.

\begin{equation}\label{eq_return}
    \boldsymbol{L}(s)=\boldsymbol{Y}_{\text{IBR}}(s)\,\boldsymbol{Y}_{g}(s)^{-1}.
\end{equation}

In this case study, the estimated dq-admittance model of IBR1 obtained under the measurement-noise conditions in Case 4 is used for stability assessment. Five operating points, listed in Tab.~\ref{tab:OP}, are evaluated under two grid conditions, denoted by $Z_{g1}$ and $Z_{g2}=1.3Z_{g1}$, and under different prediction-error levels. Since the estimated admittance is available at 20 discrete frequency points, the admittance samples are denoted by $\boldsymbol{Y}_{\text{IBR}}(k)$, and the corresponding eigenvalues $\lambda_1(k)$ and $\lambda_2(k)$ are computed for $k=1,\dots,20$. These frequency-indexed eigenvalues are then used to form the Nyquist eigenloci for assessing the consistency between the predicted and actual stability conclusions.

\begin{table}[h]
\caption{Operating points used for the stability study}
\label{tab:OP}
\centering
\renewcommand{\arraystretch}{1.1}
\setlength{\tabcolsep}{4pt}
\footnotesize
\begin{tabular}{c c c c c c}
\hline\hline
 & OP1 & OP2 & OP3 & OP4 & OP5 \\
\hline
$[P,Q]$ (p.u.) & $[1.0,0.0]$ & $[0.5,0.0]$ & $[0.8,0.0]$ & $[0.8,0.6]$ & $[0.8,-0.6]$ \\
\hline\hline
\end{tabular}
\end{table}

The stability prediction results are reported in Tab.~\ref{tab:case6_consistency}, where 30 stability predictions are compared with the reference results obtained from EMT time-domain simulations. The 0\% error level corresponds to the measurement-based admittance data used as the reference case. The 10\% and 13\% error levels represent the approximate admittance prediction errors obtained from the FedAvgIm framework under different measurement-noise conditions.

\begin{table}[h]
\caption{Stability prediction consistency for IBR1 under different prediction-error levels and operating points}
\label{tab:case6_consistency}
\centering
\renewcommand{\arraystretch}{1.1}
\setlength{\tabcolsep}{3.5pt}
\footnotesize
\begin{tabular}{c c c c c c c c}
\hline\hline
\textbf{Error} & \textbf{OP} 
& \multicolumn{2}{c}{$Z_{g1}$} & \textbf{Match} 
& \multicolumn{2}{c}{$Z_{g2}$} & \textbf{Match} \\
\cline{3-4} \cline{6-7}
\textbf{level} &  
& Predicted & Actual &  
& Predicted & Actual &  \\
\hline
0\%  & OP1 & Stable   & Stable & $\checkmark$ & Unstable & Unstable & $\checkmark$ \\
0\%  & OP2 & Stable   & Stable & $\checkmark$ & Unstable & Unstable & $\checkmark$ \\
0\%  & OP3 & Stable   & Stable & $\checkmark$ & Unstable & Unstable & $\checkmark$ \\
0\%  & OP4 & Stable   & Stable & $\checkmark$ & Unstable & Unstable & $\checkmark$ \\
0\%  & OP5 & Stable   & Stable & $\checkmark$ & Unstable & Unstable & $\checkmark$ \\
\hline
10\% & OP1 & Unstable & Stable & $\times$     & Unstable & Unstable & $\checkmark$ \\
10\% & OP2 & Stable   & Stable & $\checkmark$ & Unstable & Unstable & $\checkmark$ \\
10\% & OP3 & Stable   & Stable & $\checkmark$ & Unstable & Unstable & $\checkmark$ \\
10\% & OP4 & Stable   & Stable & $\checkmark$ & Stable   & Unstable & $\times$ \\
10\% & OP5 & Stable   & Stable & $\checkmark$ & Unstable & Unstable & $\checkmark$ \\
\hline
13\% & OP1 & Stable   & Stable & $\checkmark$ & Unstable & Unstable & $\checkmark$ \\
13\% & OP2 & Stable   & Stable & $\checkmark$ & Unstable & Unstable & $\checkmark$ \\
13\% & OP3 & Stable   & Stable & $\checkmark$ & Unstable & Unstable & $\checkmark$ \\
13\% & OP4 & Stable   & Stable & $\checkmark$ & Stable   & Unstable & $\times$ \\
13\% & OP5 & Stable   & Stable & $\checkmark$ & Unstable & Unstable & $\checkmark$ \\
\hline\hline
\end{tabular}
\end{table}

Under different operating points and grid conditions, the measurement-based admittance provides fully consistent stability predictions in the 0\% error case, achieving 100\% agreement with the EMT reference results. When admittance models with approximately 10\% prediction error are used, two out of ten stability decisions are inconsistent with the EMT reference, resulting in an 80\% agreement rate. When the prediction error increases to approximately 13\%, only one out of ten stability decisions is inconsistent, corresponding to a 90\% agreement rate. Overall, 27 out of the 30 stability predictions are consistent with the EMT reference results.

\begin{figure}[h]
\centerline{\includegraphics[scale = 1]{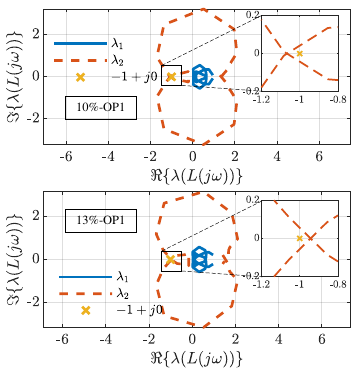}}
\caption{Generalized Nyquist eigenloci of the return-ratio matrix for OP1 under two prediction-error levels: $10\%$ error (top) and $13\%$ error (bottom).}
\label{fig:Case6_OP1}
\end{figure}

The non-monotonic relationship between admittance prediction error and stability-prediction accuracy is mainly due to the location of the generalized Nyquist eigenloci relative to the critical point $(-1+j0)$. For some operating points and grid conditions, the eigenloci are very close to the critical point. As a result, small admittance-estimation errors can shift the eigenloci across the critical point and change the predicted stability conclusion. This behavior is illustrated in Fig.~\ref{fig:Case6_OP1}, where the 10\% prediction-error case leads to an incorrect encirclement of the critical point, whereas the 13\% prediction-error case remains consistent with the EMT reference. Therefore, the impact of admittance-estimation error on stability assessment is not strictly monotonic, but depends on how the error perturbs the eigenloci relative to the critical point.

\begin{figure}[h]
\centerline{\includegraphics[scale = 1]{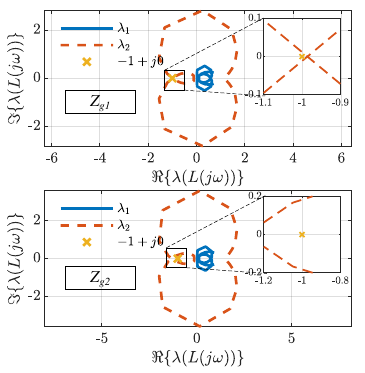}}
\caption{Nyquist plots of the eigenvalues of the return-ratio matrix under two grid impedances: $Z_{g1}$ (top) and $Z_{g2}$ (bottom).}
\label{fig:Case6_Nyquist}
\end{figure}

Fig.~\ref{fig:Case6_Nyquist} shows the generalized Nyquist eigenloci derived from the estimated admittance of IBR1 at OP1 with a 13\% prediction error. The two eigenvalues of the return-ratio matrix are evaluated under two grid-impedance scenarios: $Z_{g1}$ and $Z_{g2}$, where $Z_{g2}=1.3Z_{g1}$ represents a weaker grid condition. Under $Z_{g1}$, the Nyquist eigenloci do not encircle the critical point $(-1+j0)$, indicating stable operation. Under the weaker grid condition $Z_{g2}$, the Nyquist eigenlocus of $\lambda_2$ encircles the critical point, indicating instability.

\begin{figure}[h]
\centerline{\includegraphics[scale = 1]{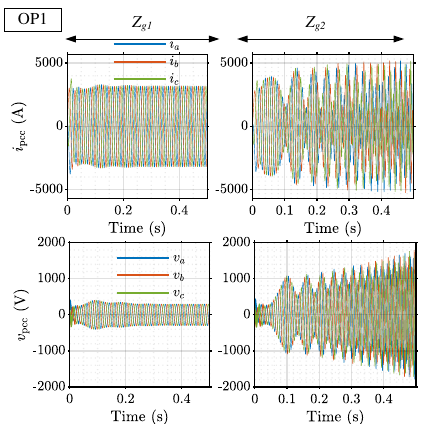}}
\caption{EMT time-domain responses of the three-phase PCC current $i_{\text{pcc}}$ (top) and PCC voltage $v_{\text{pcc}}$ (bottom) under two grid-impedance scenarios, $Z_{g1}$ and $Z_{g2}$.}
\label{fig:Case6_time}
\end{figure}

To validate this conclusion, two EMT time-domain simulations are performed separately in MATLAB/Simulink for the two grid-impedance scenarios using the same configuration as in Fig.~\ref{fig_GFLIConfig}. The PCC voltage and current responses are shown in Fig.~\ref{fig:Case6_time}. Under $Z_{g1}$, the PCC voltage and current responses remain bounded, confirming stable operation. In contrast, under the weaker-grid scenario \(Z_{g2}\), the PCC voltage exhibits rapidly growing oscillations, indicating instability. These EMT results are consistent with the generalized Nyquist assessment based on the estimated admittance model of IBR1.

\section{Conclusion}
FedAvgIm is presented as a federated-learning framework for multi-site identification of IBR small-signal behavior in the dq domain, using dq-admittance as the direct learning target. By keeping EMT measurement-based operating-point and frequency-response data local to each inverter site and exchanging only model parameters, the framework reduces direct raw-data exposure while addressing severe local data scarcity. A lightweight feed-forward neural network is trained across nine grid-following inverters to map $(V,P,Q,f)$ to the stacked real and imaginary parts of the dq-admittance entries.

The evaluation results show that FedAvgIm achieves performance close to centralized training and consistently outperforms local-only training, especially in data-limited cases where isolated local models tend to overfit. The client-joining and transfer-learning studies further show that federation growth and federated multi-source pretraining can improve generalization under raw-data-locality constraints. FedAvgIm also maintains stable learning behavior under different measurement-noise levels. Under mixed IBR structures, it remains beneficial compared with local-only learning, although its advantage decreases under severe topology and parameter heterogeneity due to client drift. The estimated dq-admittance models are further applied to impedance-based stability assessment through generalized Nyquist analysis and EMT time-domain simulations across multiple operating points and grid-strength conditions.

However, several limitations remain as follows.
\begin{itemize}
    \item First, federated learning does not inherently provide formal privacy guarantees; in this work, the privacy-related benefit is limited to reducing direct raw-data exposure and avoiding centralized raw-data pooling.
    \item Second, when sufficient local data are available, local-only training or transfer learning may achieve comparable performance.
    \item \rev{Third, a single global model may not generalize well across highly heterogeneous IBR fleets; clustered or personalized federated learning should be investigated for broader heterogeneity.}
    \item \rev{Fourth, admittance-estimation errors can affect the final stable/unstable decision, especially under weak-grid or near-critical conditions where the generalized Nyquist eigenloci are close to the critical point, so formal error bounds and uncertainty quantification are important future directions.}
    \item \rev{Fifth, the present validation is based on EMT measurement-based admittance datasets rather than field or HIL measurements. Future work will further evaluate FedAvgIm using HIL platforms or field-measured IBR data.}
    \item Finally, although this study focuses on grid-following inverters, future work will extend the framework to grid-forming inverters and mixed GFL--GFM systems.
\end{itemize}

\section*{Acknowledgment}
This work was supported in part by the U.S. National Science Foundation (NSF) under Award 2509993 \rev{and 2552448}.

\bibliographystyle{ieeetr}
\bibliography{Ref}

\end{document}